\documentclass[twocolumn,showpacs,preprintnumbers,amsmath,amssymb]{revtex4}
\usepackage{graphicx}
\usepackage{epsfig}
\usepackage{color}

\begin{document}

\title{Energy entanglement in normal metal-superconducting forks}

\author{K.V. Bayandin}
\affiliation{Landau Institute for Theoretical Physics RAS, 117940 Moscow, Russia}
\affiliation{Centre de Physique Th\'eorique, Case 907 Luminy, 13288 Marseille Cedex 9, France}
\author{G.B. Lesovik}
\affiliation{Landau Institute for Theoretical Physics RAS, 117940 Moscow, Russia}
\author{T. Martin}
\affiliation{Centre de Physique Th\'eorique, Case 907 Luminy, 13288 Marseille Cedex 9, France}
\affiliation{Universit\'e de la M\'edit\'erann\'ee, 13288 Marseille Cedex 9, France}


\begin{abstract}
The possibility for detecting energy entanglement in normal
metal--superconductor junctions is examined. For the first time we
proved that two electrons in a NS structure originating from the
same Cooper pair are entangled in the energy subspace. This work
follows previous works where spin entanglement was studied in
similar circuits. The device consists of a superconducting beam
splitter connected to two electronic Mach-Zehnder interferometers.
In each arms of the interferometers, energies are filtered with
coherent quantum dots. In contrast to previous studies of
zero-frequency cross-correlations of electrical currents for this
system, attention is drawn to finite-time measurements. This
allows to observe two-particle interference for particles with
different energies above and below Fermi level. Entanglement is
first characterized via the concurrence for the two-particle
spatial density matrix. Next, we formulate the Bell inequality,
which is written in terms of finite-time noise correlators, and
thereby we find a specific set of parameters for which
entanglement can be detected.
\end{abstract}


\maketitle

\section{Introduction}

Entanglement is the building block of quantum information
processing schemes~\cite{bouw}. In recent years, it has generated
considerable excitement in the mesoscopic physics and nanophysics
community. Over the years, electronic analogs of optical setups
have been conceived in order to probe entanglement between
electrons. In the context of electron transport, entanglement was
first described
theoretically~\cite{lesovik_martin_blatter,recher_sukhorukov_loss}
in forked normal metal-superconductor devices revealing positive
noise cross-correlations~\cite{torres_martin}. Then entanglement
in such devices was quantified via the Bell inequality
test~\cite{chtchelkatchev}.

In all these schemes a superconductor injects Cooper pairs in two
metallic arms, or in quantum dots connected to leads. When two
electrons are emitted from the superconductor, they either get
into the same lead or in opposite leads. The resulting zero
frequency noise correlations between the two normal arms will have
a tendency to be negative in the former process due to the
partition noise of Cooper pairs. The latter process, also called
crossed-Andreev process, is the source of entanglement, and leads
to positive noise correlations. Indeed, the Cooper pair is
transmitted as a singlet state, and it is entangled both in spin
and in energy.

\begin{figure}
\centerline{\includegraphics[width=7.cm]{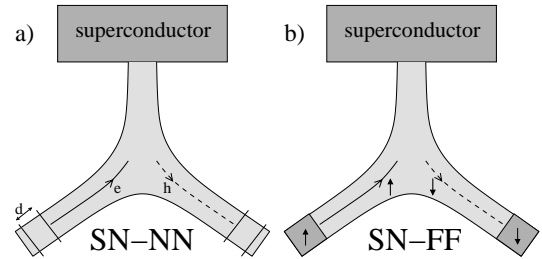}} \caption{a)
Normal metal-superconducting fork with energy filters selecting
electron ($+E$) on the left side and selecting holes ($-E$) on the
right side. b) Same fork, but this time with spin filters, leaving
the possibility of energy entanglement as both electrons and holes
(with a definite spin for each side) can propagate in both
branches.} \label{forks1}
\end{figure}

The detection of spin entanglement has been addressed on several
occasions for ideal devices~\cite{burkard_loss,saraga}, and with
the assumption that parasitic effects are
present~\cite{sauret_martin_feinberg}. In optics, a standard test
for studying the degree of entanglement of two photons is to
perform a Bell/Clauser Horne inequality test, in a coincidence
measurement. This test allows to probe the non-local nature of
quantum mechanics: are the two particles in an entangled state? In
contrast, in condensed matter physics, experiments are often
performed in a stationary state, with constant currents flowing at
the output. Nevertheless, in Ref.~\cite{chtchelkatchev} it was
shown that a Bell inequality test~\cite{kawabata} can be expressed
in terms of zero frequency noise cross-correlations at the output.
Subsequently, similar Bell inequality tests were proposed for
orbital entanglement in a superconducting
setup~\cite{orbital_entanglement}, in normal metal
devices~\cite{Entanglement3,Entanglement4}, and in Quantum Hall
systems~\cite{beenakker,buttiker}.

Ref.~\cite{lesovik_martin_blatter} proposed that some constraints
should be imposed on a plain superconducting-normal metal
fork~\cite{torres_martin} in order to probe entanglement. Energy
filters where added on each side [Fig.~\ref{forks1}(a)], selecting
positive/negative energies depending on the side. An incoming hole
thus could only be reflected as an electron in the opposite
branch, yet the spin degree of freedom was untouched. Energy
entanglement was also proposed~\cite{lesovik_martin_blatter} using
spin filters instead of energy filters [Fig.~\ref{forks1}(b)].
Except for the characterization of positive noise correlations
which constitutes a symptom of entanglement, so far no
quantitative test of this entanglement has been reached.

Recent advances in quantum optic experiments on momentum-phase
entanglement~\cite{rarity} led to experimental realization of such
techniques as two-particle teleportation~\cite{teleportation}, the
purification protocol for polarization entangled
state~\cite{purification} and the full Bell-state
analysis~\cite{bell_analiser}. All of the above provide additional
motivation for us to reexamine the question of energy
entanglement.

The purpose of this work is to explain how energy entanglement can
be detected in normal metal-superconducting forks. Although the
setup which we propose is built from elementary building blocks of
mesoscopic physics, it is rather elaborate. We thus emphasize that
our main goal is to show whether energy entanglement detection is
plausible from a theoretical standpoint.

The paper is organized as follow: Sec. 2 gives a basic description
of the setup, which is composed of a superconducting fork
connected to two Mach-Zehnder (MZ) interferometers. Sec. 3 and 5
describe the calculation of the two-particle density matrix and
electrical current cross-correlations in the context of scattering
theory for this normal metal-superconductor setup. Sec. 4
describes the entanglement characterization via the concurrence of
the density matrix. Sec. 6 develops the Bell inequality analysis
in terms of electrical currents.

\section{Description of the setup}

\begin{figure}
\centerline{\includegraphics[width=7.cm]{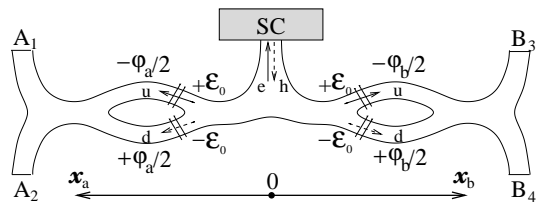}}
\caption{Setup for measuring energy entanglement from a normal
metal-superconducting fork. The superconductor emits Cooper pairs
in the ballistic region. The electrons from a pair are then spit
on the first fork. Then the particles are filtered according to
their energy. The amplitudes for particles with energies above and
below Fermi level are combined at the last beam splitter, in close
analogy with the optical setup~\cite{rarity}.} \label{fig2}
\end{figure}

We now present a setup (Fig.~\ref{fig2}), which is able to test
energy entanglement via the violation of Bell inequalities. We use
a normal-metal fork connected to a superconductor. The leads
connected to the arms of this fork are assumed to have
quasiparticle mean free path larger than the size of the setup: in
this ideal version of the setup, backscattering is absent. A
possibility would be to use electrostatically defined quantum
wires on semiconductor heterostructures, assuming that the
interface between the two dimensional electron gas and the
superconductor can be controlled. The two leads connected to the
injection fork (center of Fig.~\ref{fig2}) are each attached to
two other forks (left and right of Fig.~\ref{fig2}), which filter
the particles according to their positive/negative energy,
measured with respect to the chemical potential of the
superconductor. This energy filtering can in principle be realized
via a double barrier structure forming an electronic Fabry-Perot
interferometer. An example of such a filter was demonstrated in a
"which path detection" experiment at the Weizmann
Institute~\cite{Umansky}) where one arm of the devices contained a
coherent quantum dot.

Next, the two forks on the right and on the left are reconnected,
forming MZ interferometers. The paths on each arm of such
interferometers have equal length, and the two beam splitters are
assumed to be symmetrical. Later on in this paper, we shall
consider the measurements of current cross-correlations performed
between different contacts $A_{1}$, $A_{2}$, $B_{3}$ and $B_{4}$
of these two interferometers. Note that, in this suggested
geometry of the setup, there is no absolute need for spin-filters,
contrarily to what was suggested in
Ref.~\onlinecite{lesovik_martin_blatter} and illustrated in
Fig.~\ref{forks1}(b). As we shall see below, if the interface
between the injecting fork and the superconductor is opaque, the
two constituent electrons of a Cooper pair are split between the
right and left side and the cross correlation signal is positive,
as in a photon coincidence measurement.

The idea of this experiment is that two electrons, which originate
from a Cooper pair in the superconductor, once on the normal metal
side, have energies which are symmetric with respect to the
chemical potential of the superconductor. In principle they can be
filtered in a way that an electron with energy above this Fermi
level always gets into the upper arms of the interferometers
(energy $+\epsilon_{0}$, leads $u_{A}$ and $u_{B}$ in
Fig.~\ref{fig2}) or it gets into the lower arms of interferometers
otherwise (energy $-\epsilon_{0}$, leads $d_{A}$ and $d_{B}$ in
Fig.~\ref{fig2}). That is, two electrons being split at the
injection superconducting  fork (center of Fig.~\ref{fig2}) always
end up into the right upper and the left lower arms of the
interferometers or into the right lower and the left upper ones.

This situation is very similar to the phase-momentum entanglement
in quantum optics. Indeed, an experimental non-locality test for
photons was achieved by Rarity and Tapster using two MZ
interferometers~\cite{rarity}. Two photons with different wave
lengths were generated during a down-conversion process, and
subsequent recombination of the beams allowed to perform a
correlation measurement. We find that this quantum optics
experiment can be translated for and applied to electronic
devices. As we shall see, in our condensed matter analog, which
uses charged particles, the Bell inequality violation can be tuned
by varying the magnetic fluxes within interferometers (electrons
gain extra phases $e^{\pm i\varphi_{a}/2}$ and $e^{\pm
i\varphi_{b}/2}$ in the lower (upper) arms of the
interferometers).

In the quantum optics experiment~\cite{rarity}, the momentum
direction of the down-converted photons allowed to separate the
photons and to let them recombine at the beam splitters, provided
that a single beam splitter can be reached only by photons with
equal frequencies. This property is essential for the two-particle
interference. In our proposal the particles, which are reaching
the last beam splitters (recombination of the beams) along
different paths, have different energies, so it is in principle
possible to achieve "which path" information and therefore to
destroy the interference. We propose to use measurements on {\it
short enough time scales so that it is not possible to resolve the
energy of the particle} (energy/time Heisenberg principle). We
claim that the two-particle interference and the non-local
properties of quantum mechanics can be detected in this ideal
setup.

\section{Calculation of the density matrix}

In order to analyze the properties of the injected electrons, we
calculate the two-particle density matrix for quasiparticles
caught right after energy filters (in leads $u_{A}$, $d_{A}$ and
$u_{B}$, $d_{B}$):
\begin{equation}
g(\vec{\sigma},\vec{x})=Tr\left[\hat\rho
\hat\Psi_{\sigma_{1}^{\prime}}^{\dagger}(x_{i^{\prime}})\hat\Psi_{\sigma_{2}^{\prime}}^{\dagger}(x_{j^{\prime}})
\hat\Psi_{\sigma_{2}}(x_j)\hat\Psi_{\sigma_{1}}(x_i)\right],\label{DM}
\end{equation}
here we have written the whole matrix with non-diagonal elements,
and the indexes $\sigma$ describe the spins part of density
matrix, and orbital indexes $i=u_{A},d_{A}$ and $j=u_{B},d_{B}$
describe the lead there the particle is caught. In practice, only
diagonal elements with $x_{i}=x_{i^{\prime}}$ and
$x_{j}=x_{j^{\prime}}$ can be measured, but our task is to analyze
the amount of entanglement which is implicit in such a matrix,
therefore non-diagonal elements play a crucial role.

In our calculations we follow the scattering theory approach
developed in~\cite{Noise1,Noise2}, and later applied to NS-systems
with Andreev scattering~\cite{Noise_NS}. For background purposes,
the reader is referred to the
reviews~\cite{Blanter_Buttiker,Martin_Review}. Here we follow
closely the notations of Ref.~\onlinecite{Noise3}. For simplicity,
we consider ballistic quasi one-dimensional quantum wires without
backscattering. The transport in this system is governed by the
properties of Andreev reflection and normal reflection of
electrons and holes. So it is useful to perform Bogoliubov
transformation (see Ref.~\onlinecite{bogolubov_de_gennes}) for the
annihilation operator of a particle  at a position $x$ with
up/down spin $\sigma=\pm 1$:
\begin{eqnarray}
&&\hat{\Psi}_{i,\sigma}(x_{i},t_{i})=\frac{1}{\sqrt{2\pi}}\sum_{j,\beta}\int_{0}^{+\infty}dE
\left(\frac{u_{ij\beta}(x_{i})}{\sqrt{\hbar
v_{e}^{j}(E)}}\hat{c}_{j\beta\sigma}(E,t_{i})-\right.\nonumber\\
&&~~~~~~~~~~~~~~~~
\left.-\sigma\frac{v_{ij\beta}^{*}(x_{i})}{\sqrt{\hbar
v_{h}^{j}(E)}}\hat{c}_{j\beta-\sigma}^{+}(E,t_{i})\right).\label{WF}
\end{eqnarray}

The states $u_{ij\beta}$ or $v_{ij\beta}$ correspond to wave
functions of an electron or a hole scattered in lead $i$, due to a
quasiparticle of type $\beta=e,h$ incoming from lead $j$.
Operators $\hat{c}_{j\beta\sigma}(E)$ and
$\hat{c}_{j\beta\sigma}^{+}(E)$ satisfy standard Fermi statistic
relations. $v_{e}^{j}(E)=\hbar k_{e}^{j}(E)$ and
$v_{h}^{j}(E)=\hbar k_{h}^{j}(E)$ are velocities of electrons and
holes in lead $j$.

The evolution of particle states $u_{ij\beta}$ and  $v_{ij\beta}$
is described by Bololiubov-de Gennes equations, which can be
written as
\begin{subequations}
\begin{eqnarray}
&&\left(-\frac{\hbar^2}{2m}\frac{\partial^2}{\partial x^2}
-\mu_{S}+V(x)\right)u_{ij\beta}(x)\nonumber\\
&&~~~~~~~~~~~~~~~~~~~~
+\Delta(x)v_{ij\beta}(x)=Eu_{ij\beta}(x),\label{BdG_1}\\
&&\left(+\frac{\hbar^2}{2m}\frac{\partial^2}{\partial x^2}
+\mu_{S}-V(x)\right)v_{ij\beta}(x)\nonumber\\
&&~~~~~~~~~~~~~~~~~~~~
+\Delta^{*}(x)u_{ij\beta}(x)=Ev_{ij\beta}(x).\label{BdG_2}
\end{eqnarray}
\end{subequations}
The pair potential $\Delta(x)$ should be
calculated selfconsistently, but, for simplicity, in our
calculation it corresponds to the superconducting gap $\Delta$ in
the bulk of the superconductor ($x>0$), and it is zero in the
normal leads ($x<0$).

In a normal ideal lead $\Delta(x)=0$ and $V(x)=0$, Bogoliubov-de
Gennes equations~(\ref{BdG_1})-(\ref{BdG_2}) reduce to Schrodinger
equation for free electrons and holes. Solutions of form
$e^{ik_{e}^{N}x}$ for electrons and $e^{ik_{h}^{N}x}$ for holes
are chosen, where $k_{e}^{N}=\sqrt{2m(\mu_{S}+E)}/\hbar$ and
$k_{h}^{N}=\sqrt{2m(\mu_{S}-E)}/\hbar$ are wave vectors for
electrons and holes.

Electrons and holes that originate from a particle of type $\beta$
in lead $j$ and scattered into lead $i$ are described by:
\begin{subequations}
\begin{eqnarray}
&&u_{ij\beta}(x_{i})=\delta_{ij}\delta_{e\beta}e^{+ik_{e}^{N}x_{i}}+
S_{ij}^{e\beta}\sqrt{\frac{k_{\beta}^{j}}{k_{e}^{N}}}e^{-ik_{e}^{N}x_{i}},\label{WF_u}\\
&&v_{ij\beta}(x_{i})=\delta_{ij}\delta_{h\beta}e^{-ik_{h}^{N}x_{i}}+
S_{ij}^{h\beta}\sqrt{\frac{k_{\beta}^{j}}{k_{h}^{N}}}e^{+ik_{h}^{N}x_{i}}.\label{WF_v}
\end{eqnarray}
\end{subequations}
Here we have the opposite sign for momenta of
electrons and holes, which is due to the symmetry of Bogoliubov-de
Gennes equations. For simplicity, we introduced a notation
$S_{ij}^{\alpha\beta}$ for scattering-matrix element expressing
the amplitude of an outgoing particle of type $\alpha$ in lead $i$
due to an incident particle of type $\beta$ in lead $j$. In
general, these amplitudes depend on the  quasiparticle energies.

\begin{widetext}
We linearize the wave vectors in energy in the exponents, and then
take them equal to $k_{F}$ in pre-exponential factors. We
substitute all the equations into Eq. (\ref{DM}) and obtain:
\begin{eqnarray}
&&\;\;\;\;\;\;\;\;g(\vec{\sigma},\vec{x})=\frac{1}{(hv_{F})^{2}}\sum\limits_{k,l,m,n}\int_{0}^{+\infty}\dots\int_{0}^{+\infty}dE_{1}\dots dE_{4}\times\label{for_appendix1}\\
&&\left[\left\langle\hat{c}_{k\sigma_{1}^{\prime}}^{\dagger}(E_{1},t_{i^{\prime}})\hat{c}_{l\sigma_{2}^{\prime}}^{\dagger}(E_{2},t_{j^{\prime}})\hat{c}_{m\sigma_{2}}(E_{3},t_{j})\hat{c}_{n\sigma_{1}}(E_{4},t_{i})\right\rangle
u_{i^{\prime}k}^{*}(x_{i^{\prime}})u_{j^{\prime}l}^{*}(x_{j^{\prime}})u_{jm}(x_{j})u_{in}(x_{i})\right.\\
&&+\left\langle\hat{c}_{k\sigma_{1}^{\prime}}^{\dagger}(E_{1},t_{i^{\prime}})\hat{c}_{l-\sigma_{2}^{\prime}}(E_{2},t_{j^{\prime}})\hat{c}_{m-\sigma_{2}}^{\dagger}(E_{3},t_{j})\hat{c}_{n\sigma_{1}}(E_{4},t_{i})\right\rangle\sigma_{2}^{\prime}\sigma_{2}
u_{i^{\prime}k}^{*}(x_{i^{\prime}})v_{j^{\prime}l}(x_{j^{\prime}})v_{jm}^{*}(x_{j})u_{in}(x_{i})\\
&&+\left\langle\hat{c}_{k\sigma_{1}^{\prime}}^{\dagger}(E_{1},t_{i^{\prime}})\hat{c}_{l-\sigma_{2}^{\prime}}(E_{2},t_{j^{\prime}})\hat{c}_{m\sigma_{2}}(E_{3},t_{j})\hat{c}_{n-\sigma_{1}}^{\dagger}(E_{4},t_{i})\right\rangle\sigma_{2}^{\prime}\sigma_{1}
u_{i^{\prime}k}^{*}(x_{i^{\prime}})v_{j^{\prime}l}(x_{j^{\prime}})u_{jm}(x_{j})v_{in}^{*}(x_{i})\\
&&+\left\langle\hat{c}_{k-\sigma_{1}^{\prime}}(E_{1},t_{i^{\prime}})\hat{c}_{l\sigma_{2}^{\prime}}^{\dagger}(E_{2},t_{j^{\prime}})\hat{c}_{m-\sigma_{2}}^{\dagger}(E_{3},t_{j})\hat{c}_{n\sigma_{1}}(E_{4},t_{i})\right\rangle\sigma_{1}^{\prime}\sigma_{2}
v_{i^{\prime}k}(x_{i^{\prime}})u_{j^{\prime}l}^{*}(x_{j^{\prime}})v_{jm}^{*}(x_{j})u_{in}(x_{i})\\
&&+\left\langle\hat{c}_{k-\sigma_{1}^{\prime}}(E_{1},t_{i^{\prime}})\hat{c}_{l\sigma_{2}^{\prime}}^{\dagger}(E_{2},t_{j^{\prime}})\hat{c}_{m\sigma_{2}}(E_{3},t_{j})\hat{c}_{n-\sigma_{1}}^{\dagger}(E_{4},t_{i})\right\rangle\sigma_{1}^{\prime}\sigma_{1}
v_{i^{\prime}k}(x_{i^{\prime}})u_{j^{\prime}l}^{*}(x_{j^{\prime}})u_{jm}(x_{j})v_{in}^{*}(x_{i})\\
&&\left.+\left\langle\hat{c}_{k-\sigma_{1}^{\prime}}(E_{1},t_{i^{\prime}})\hat{c}_{l-\sigma_{2}^{\prime}}(E_{2},t_{j^{\prime}})\hat{c}_{m-\sigma_{2}}^{\dagger}(E_{3},t_{j})\hat{c}_{n-\sigma_{1}}^{\dagger}(E_{4},t_{i})\right\rangle\sigma_{1}^{\prime}\sigma_{2}^{\prime}\sigma_{2}\sigma_{1}
v_{i^{\prime}k}(x_{i^{\prime}})v_{j^{\prime}l}(x_{j^{\prime}})v_{jm}^{*}(x_{j})v_{in}^{*}(x_{i})\right],\label{for_appendix2}
\end{eqnarray}
here the summation over indexes $k$, $l$, $m$ and $n$ implies a
summation over lead indexes and types of particles - $e$ or $h$.

We consider the case where all normal leads are biased with the
same voltage $V$.
With these assumptions, the density matrix takes the form (see Appendix):
\begin{eqnarray}
&&g(\vec{\sigma},\vec{x})=\delta_{\sigma_{1}\sigma_{1}^{\prime}}\delta_{\sigma_{2}\sigma_{2}^{\prime}}G_{i^{\prime}i}G_{jj^{\prime}}^{*}-
\delta_{\sigma_{1}\sigma_{2}^{\prime}}\delta_{\sigma_{2}\sigma_{1}^{\prime}}G_{i^{\prime}j}G_{ij^{\prime}}^{*}+
I_{\sigma_{1}\sigma_{2}}I_{\sigma_{1}^{\prime}\sigma_{2}^{\prime}}F_{i^{\prime}j^{\prime}}F_{ij}^{*},
\end{eqnarray}
where we have used a notation for a matrix:
$I_{\sigma_{1}\sigma_{2}}=i\hat{\tau}_y$ and $\hat{\tau}_y$ is a
Pauli matrix. We compute this density matrix  at locations taken
in the leads of the interferometers: $u_{A}$, $d_{A}$, $u_{B}$ and
$d_{B}$ (See Fig.~\ref{fig2}). We thus define indices $i$ and
$i^{\prime}$ which can take the values  $u_{A}$ or $d_{A}$ and
indexes $j$ and $j^{\prime}$ equal to $u_{B}$ or $d_{B}$. We also
use a notation for the two-point correlation functions:
\begin{subequations}
\begin{eqnarray}
&&G_{ij}(x_{i},x_{j},t_{j}-t_{i})=\frac{1}{hv_{F}}\sum\limits_{k}\int_{-\infty}^{+\infty}dE f_{k}(E)(u_{ik}^{*}(x_{i})u_{jk}(x_{j}))e^{-iE(t_{j}-t_{i})},\label{G_corr}\\
&&F_{ij}(x_{i},x_{j},t_{j}-t_{i})=\frac{1}{hv_{F}}\sum\limits_{k}\int_{-\infty}^{+\infty}dE
f_{k}(E)(u_{ik}^{*}(x_{i})v_{jk}(x_{j}))e^{-iE(t_{j}-t_{i})}.\label{F_corr}
\end{eqnarray}
\end{subequations}
\end{widetext}

Up to now, we have not specified the transmission properties of
the combined normal metal fork/superconductor plus energy
filtering forks. For the parametrization of the beam splitter
connected to the superconductor we take a scattering matrix with
real amplitudes described in~\cite{gefen_buttiker}:
\begin{equation}
U_{SL_{0}R_{0}}=\left(
\begin{array}{ccc}
 -(a+b) & \sqrt{\epsilon} & \sqrt{\epsilon}\\
 \sqrt{\epsilon} & a & b\\
 \sqrt{\epsilon} & b & a
\end{array}
\right),\label{epsilon_param}
\end{equation}
with the free parameter $0 \le \epsilon\le \frac{1}{2}$ describing
the transmission probability between the top lead and the right
and left arms. In general, there are two possible choices of
parametrization~\cite{lesovik_martin_blatter}:
$a=\frac{\sqrt{1-2\epsilon}-1}{2}$,
$b=\frac{\sqrt{1-2\epsilon}+1}{2}$, or alternatively $a$ and $b$
can be exchanged. Here we use the first parametrization: it
corresponds to a good transmission for electrons (holes) entering
from the left lead of the fork and exiting from the right one, and
vice versa. We claim that the excess parts of all the correlators
are the same if one choses the opposite convention. This unitary
matrix (\ref{epsilon_param}) is assumed to be the same for
electrons and holes, which implies a weak dependence of the
scattering matrix on energy. Due to multiple Andreev reflection on
$NS$-boundary and a phase shift in the superconducting lead (S),
the transmission amplitudes change in an analogous manner to a
Fabry-Perot interferometer. Assuming ideal Andreev reflection on
the boundary of the superconductor:
$r_{SS}^{eh}=r_{SS}^{he}=e^{-i\arccos{\frac{E}{\Delta}}}$, and
$e^{-i\eta}=r_{SS}^{eh}e^{-iE\frac{2S}{v_{F}\hbar}}$ one can get
\begin{subequations}
\begin{eqnarray}
&&A=r_{LL}^{\alpha\alpha}=r_{RR}^{\alpha\alpha}=
a-\frac{\epsilon(a+b)e^{-2i\eta}}{1-(a+b)^{2}e^{-2i\eta}},\label{ABCD1}\\
&&B=t_{LR}^{\alpha\alpha}=t_{RL}^{\alpha\alpha}=
b-\frac{\epsilon(a+b)e^{-2i\eta}}{1-(a+b)^{2}e^{-2i\eta}},\\
&&C=r_{LL}^{\alpha\beta}=r_{RR}^{\alpha\beta}=
\frac{\epsilon e^{-i\eta}}{1-(a+b)^{2}e^{-2i\eta}},\\
&&D=t_{LR}^{\alpha\beta}=t_{RL}^{\alpha\beta}= \frac{\epsilon
e^{-i\eta}}{1-(a+b)^{2}e^{-2i\eta}},\label{ABCD2}
\end{eqnarray}
\end{subequations} here $A$ and $B$ are reflection and transmission
amplitudes for particles of the same type ($\alpha\alpha$) with
incident and final lead to the left ($L$) and to the right ($R$)
from the fork, and $C$ and $D$ are reflection and transmission
amplitudes for the case of changed types of particles
($\alpha\beta$). Although notations $A$, $B$, $C$, $D$ are not
used here, they are introduced for later convenience, in Eqs.
(\ref{s_matrx1}) and (\ref{s_matrx2}).

The energy filtering forks are chosen ``ideal'': electrons with
energies above the superconducting chemical potential pass without
reflection through the upper arm, while electrons which energy is
below this chemical potential go in the lower arm of the fork
(left side of Fig.~\ref{fig2}). The following transmission
amplitudes are thus chosen for the left energy filtering fork:
\begin{subequations}
\begin{eqnarray}
&r_{LL}^{ee}(\pm E)=0,&\\
&r_{uu}^{ee}(+E)=0,&\;\;\;\;r_{uu}^{ee}(-E)=-1,\label{udL1}\\
&r_{dd}^{ee}(+E)=+1,&\;\;\;\;r_{dd}^{ee}(-E)=0,\\
&t_{uL}^{ee}(+E)=1,&\;\;\;\;t_{uL}^{ee}(-E)=0,\\
&t_{dL}^{ee}(+E)=0,&\;\;\;\;t_{uL}^{ee}(-E)=-i, \label{udL2}
\end{eqnarray}
\end{subequations}
here the amplitudes for hole type quasiparticles
follow from Eq. (\ref{eh_symm}) and the same is correct for the
right fork exchanging $L$ on $R$. This is a very simple
parametrization, and the only constraint is the unitary
conditions. In a real experiment there may be a more complicated
dependence on energy leading to imperfections, but this question
goes beyond the scope of the paper.

Now the global scattering amplitudes are calculated in a simple
way, because of our assumption of no backscattering in the leads.
For example:
\begin{equation}
S_{u_{L}d_{L}}^{ee}=r_{uu}^{ee}+t_{uL}^{ee}S_{LL}^{ee}t_{Lu}^{ee}~.
\end{equation}

These amplitudes allow to calculate the pair correlation functions
from the Eqs. (\ref{G_corr}) and (\ref{F_corr}), and the results
may be divided into an equilibrium and an excess part:
\begin{equation}
G_{ij}(x_{i},x_{j},\Delta t_{ij})=G_{ij}^{eq}(x_{ij}^{+},\Delta
t_{ij})+G_{ij}^{ex}(\Delta x_{ij},\Delta t_{ij})~,
\end{equation}
the former depends on the sum of coordinates and describes the
Fermi correlations within normal leads, the latter depends on the
difference of coordinates and describes the particles injected
from the superconductor. The most interesting contributions for
the  discussion below are the excess contributions which dominate
at low temperatures ($T\ll eV$):
\begin{subequations}
\begin{eqnarray}
\small
&G_{ij}^{ex}=&-\delta_{ij}\frac{\epsilon^{2}}{2(1-\epsilon)^{2}}e^{-ik_{F}\Delta x_{ij}}e^{\pm i\frac{eV}{2v_{F}\hbar}(\Delta x_{ij}+\Delta t_{ij})}\nonumber\\
&&\times\frac{\sin{\frac{eV}{2v_{F}\hbar}(\Delta x_{ij}+\Delta t_{ij})}}{\pi(\Delta x_{ij}+\Delta t_{ij})},\\
&F_{ij}^{ex}=&\delta_{i\overline{j}}\frac{\epsilon\sqrt{1-2\epsilon}}{2(1-\epsilon)^{2}}e^{+ik_{F}x_{ij}^{+}}e^{\pm i\frac{eV}{2v_{F}\hbar}(\Delta x_{ij}+\Delta t_{ij})}\nonumber\\
&&\times\frac{\sin{\frac{eV}{2v_{F}\hbar}(\Delta x_{ij}+\Delta
t_{ij})}}{\pi (\Delta x_{ij}+\Delta t_{ij})},
\label{density_result}
\end{eqnarray}
\end{subequations}
where $i=u_{A},d_{A}$ and $i=u_{B},d_{B}$; $\epsilon$ is the
scattering parameter of Eq.(\ref{epsilon_param}) of the
superconducting fork; $\Delta x_{ij}=x_{j}-x_{i}$ and $\Delta
t_{ij}=t_{j}-t_{j}$; and $\delta_{ij}$ ($\delta_{i\overline{j}}$)
is equal to unity (zero) only if $i$ and $j$ are simultaneously
$u$ or $d$.

\section{Concurrence for the density matrix}

We now turn to the calculation of the concurrence, as defined in
the work of Wootters~\cite{wootters}, but adapted to our transport
geometry. With the additional assumption of simultaneous
measurements (all the times $t_{i}$ in Eq. (\ref{density_result})
are the same) and equal coordinates on the same sides of the setup
($x_{u_{A}}=x_{d_{A}}$ and $x_{u_{B}}=x_{d_{B}}$, up to a the
precision $\delta x\ll v_{F}\tau_{V}$) the whole density matrix
simplifies to:
\begin{eqnarray}
&g(\vec{\sigma},\vec{x})=&g_{0}^{2}\left[\delta_{\sigma_{1}\sigma_{1}^{\prime}}\delta_{\sigma_{2}\sigma_{2}^{\prime}}\delta_{i^{\prime}i}\delta_{jj^{\prime}}\nonumber\right.\\
&&-\delta_{\sigma_{1}\sigma_{2}^{\prime}}\delta_{\sigma_{2}\sigma_{1}^{\prime}}\delta_{i^{\prime}j}\delta_{ij^{\prime}}e^{\pm i\varphi}g^{2}(\Delta_ x)\nonumber\\
&&\left.+I_{\sigma_{1}\sigma_{2}}I_{\sigma_{1}^{\prime}\sigma_{2}^{\prime}}\delta_{i^{\prime}\overline{j^{\prime}}}\delta_{i\overline{j}}e^{\pm
i\varphi}f^{2}(\Delta_ x)\right],\label{SS_DM}
\end{eqnarray}
where
\begin{equation}
g_{0}=\frac{\epsilon^{2}}{2(1-\epsilon)^{2}}\frac{eV}{2\pi
v_{F}\hbar},\;\;\;\; g(\Delta x)=\frac{\sin{\frac{\Delta
x}{2v_{F}\tau_{V}}}}{\frac{\Delta x}{2v_{F}\tau_{V}}},
\end{equation}
\begin{equation}
f(\Delta x)=\frac{\sqrt{1-2\epsilon}}{\epsilon}g(\Delta x),
\end{equation}
where $\tau_{V}=\hbar/eV$. For $i=j^{\prime}=u$ and
$i^{\prime}=j=d$, the parameter
$\varphi=\frac{eV}{v_{F}\hbar}\Delta x$ which appear in Eq.
(\ref{SS_DM}) is taken with a plus sign, for $i=j^{\prime}=d$ and
$i^{\prime}=j=u$ - with a minus sign, and in all other cases it is
zero.

The two-particle density matrix describing the spin entangled
state of two quasi-particles inside a superconductor has been
already studied in~\cite{Kim} for a bulk superconductor. In
contrast, our analysis allows to study energy entanglement in a
condensed matter setting with normal metal leads, which are
essential for the diagnosis of entanglement. Since we are not
interested in the spin entanglement here, we trace Eq.
(\ref{SS_DM}) over spin degrees of freedom, and we obtain a
density matrix which depends only on orbital degrees of freedom:
\begin{equation}
g_{i,j;i^{\prime},j^{\prime}}=Tr_{\sigma}g(\vec{\sigma},\vec{x})=2g_{0}^{2}N\hat\rho_{AB}
\end{equation}
\begin{equation}
\hat\rho_{AB}=\frac{1}{N}\left(
\begin{array}{cccc}
  2-g^{2} & 0 & 0 & 0 \\
  0 & 2+f^{2} & (f^{2}-g^{2})e^{+i\varphi} & 0 \\
  0 & (f^{2}-g^{2})e^{-i\varphi} & 2+f^{2} & 0 \\
  0 & 0 & 0 & 2-g^{2}
\end{array}\right),
\end{equation}
where the normalization factor $N=8+2f^{2}-2g^{2}$ is chosen so
that the trace for this matrix is unity. The parameter $g^{2}\le
1$. It turns out that the parameter
$\varphi=\frac{eV}{v_{F}\hbar}\Delta x$ does not appear in the
final result for the entanglement measure.

One of the possible expansions (on the basis of pure states) for
such a mixed state may be written in the space of the pseudo-spin
$|\uparrow\rangle=|u\rangle$ and $|\downarrow\rangle=|d\rangle$:
\begin{eqnarray}
&&\hat\rho_{AB}=\frac{1}{N}\left[(f^{2}-g^{2})|\Psi^{(+)}\rangle\langle\Psi^{(+)}|+(2-g^{2})\hat{\bf I}\right.\nonumber\\
&&\;\;\;\;\left.+2g^{2}\left(|\uparrow_{A}\downarrow_{B}\rangle\langle\uparrow_{A}\downarrow_{B}|+
|\downarrow_{A}\uparrow_{B}\rangle\langle\downarrow_{A}\uparrow_{B}|\right)\right],
\end{eqnarray}
where $\hat{\bf I}$ - the unity matrix, and the only entangled
term in the sum corresponds to the state:
\begin{equation}
|\Psi^{(+)}\rangle=\frac{1}{\sqrt{2}}\left(|\uparrow_{A}\downarrow_{B}\rangle+e^{+i\varphi}|\downarrow_{A}\uparrow_{B}\rangle\right).\label{psi}
\end{equation}
Note that the density matrix of Eq. (\ref{DM}) describes only a
part of the whole wave function for all electrons, and this is why
it corresponds to a mixed state. But for the case of an opaque
injection fork ($\epsilon\to 0$, $f/g\to\infty$), the state of the
orbital degrees of freedom is represented by the wave function
(\ref{psi}), which corresponds to one of the maximally entangled
states. Thus, according to the so-called "monogamy of
entanglement"~\cite{PRA_61_052306}, these degrees of freedom
cannot be entangled to any other ones, and one can say that these
particular two electrons are maximally entangled.

We now calculate the concurrence for $\hat\rho_{AB}$ - an
entanglement measure proposed by Wootters {\it et
al.}~\cite{wootters}:
\begin{equation}
C\equiv\max\{0,\sqrt{\lambda_{1}}-\sqrt{\lambda_{2}}-\sqrt{\lambda_{3}}-\sqrt{\lambda_{4}}\},
\end{equation}
where $\lambda_{1}\ge\lambda_{2}\ge\lambda_{3}\ge\lambda_{4}\ge 0$
are the eigenvalues of the matrix $\rho_{AB}\overline{\rho}_{AB}$,
and the "spin-flipped"~matrix is the folloing:
\begin{equation}
\overline{\rho}_{AB}=\hat{\sigma}_{y}^{(A)}\otimes\hat{\sigma}_{y}^{(B)}\rho^{*}_{AB}\hat{\sigma}_{y}^{(A)}\otimes\hat{\sigma}_{y}^{(B)}
\end{equation}
So for our setup the concurrence takes the form:
\begin{equation}
{\bf C}=\max{\left\{0,\frac{f^{2}-2}{4+f^{2}-g^{2}}\right\}},
\end{equation}
or using the ratio between $f$ and $g$:
\begin{equation}
{\bf
C}=\max{\left\{0,\frac{(1+\gamma)g^{2}-2\gamma}{g^{2}+4\gamma}\right\}},
\end{equation}
where $\gamma=\frac{\epsilon^{2}}{1-2\epsilon-\epsilon^{2}}$ is
the transparency parameter of the superconducting fork.

\begin{figure}
\centerline{\includegraphics[width=7.5cm,height=5.cm]{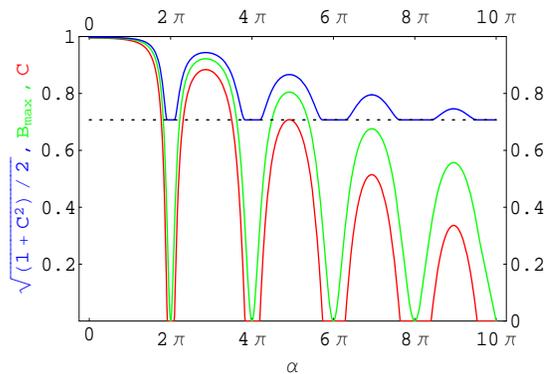}}
\caption{(Color online) Plot of the maximal Bell inequality
violation $B_{max}$ (middle green plot) as a function of the
deviation from the coincidence of the current measurements:
$\alpha=\frac{x_{B}-x_{A}}{v_{F}\tau_{V}}$, for the case of small
measurement times: $\tau\ll\hbar/eV$. $B_{max}$ is sandwiched
between the plot for concurrence ${\bf C}$ (lower red plot) for
orbital part of the two-particle density matrix: $\hat\rho_{AB}$,
and the plot for the function: $\sqrt{\frac{1+{\bf C}^{2}}{2}}$
(upper blue plot). The doted line at $1/\sqrt{2}$ marks the
critical value above which the Bell inequality is violated. The
transmission probability is taken to be:
$\gamma\approx\epsilon^{2}=0.001$.}\label{plot3}
\end{figure}

The concurrence, which depends on the coordinate difference
$\frac{\Delta x}{v_{F}\tau_{V}}$, is shown in Fig.~\ref{plot3}
choosing $\gamma=0.001$. It takes values between $0$ and $1$, the
former corresponds to absence of entanglement, and the latter to
maximal entanglement. Actually, for $\gamma$ going to zero and
$\Delta x\ll v_{F}\tau_{V}$ the concurrence is maximal, which
means that all the excess cross-correlation noise is due to the
pairs of electrons in $|\Psi^{(+)}\rangle$ maximally entangled
orbital state.

A natural question to ask is whether entanglement can occur
between orbital and spin degrees of freedom. Starting from the
general density matrix of Eq. (\ref{SS_DM}), it is possible to
perform a partial trace with respect to a single spin and a single
pseudo spin, in order to test this property. We do not show this
computation, but instead mention only the result: the concurrence
is strictly zero. In any case, if no trace of the density matrix
(\ref{SS_DM}) is performed, we conjecture that the joined state of
spin and orbital degrees of freedom can be a hyperentangled
state~\cite{PRL_72_052110}. In quantum optics these states can be
used for instance to measure the concurrence
directly~\cite{Nature_440_1022}.

\section{Calculation of current cross-correlations}

The theoretical calculation of the concurrence cannot easily be
tested experimentally: it would require quantum state tomography,
which represents a considerable challenge in experimental
mesoscopic physics. A theoretical proposal for tomography has been
been presented recently~\cite{samuelson_tomography}, which relies
on the measurement of zero frequency noise cross-correlations.
Because of the considerable experimental challenge of tomography
and because we wish to compare different diagnosis for quantum
mechanical non-locality, here we present a Bell violation test
based on the analysis of short time current cross-correlations.
Here we are interested in the same quantities for the specific
setup of Fig.~\ref{fig2}, evaluated at  points in the leads:
$A_{1}$, $A_{2}$, $B_{3}$ and $B_{4}$.

We continue using scattering theory for NS-systems. Our starting
point is the current operator in lead $i$:
\begin{equation}
\hat{I}_{i}(x)=\frac{ie\hbar}{2m}\sum_{\sigma}
\left[\frac{\partial\Psi_{i,\sigma}^{+}(x)}{\partial
x}\Psi_{i,\sigma}(x)-\Psi_{i,\sigma}^{+}(x)\frac{\partial\Psi_{i,\sigma}(x)}{\partial
x}\right],\label{CO}
\end{equation}
this operator is expressed in terms of electron and hole wave
functions:
\begin{eqnarray}
&&\hat{I}_{i}(x)=\frac{ie\hbar}{2mv_{F}}\frac{1}{2\pi\hbar}
\int_{0}^{+\infty}dE_{1}\int_{0}^{+\infty}dE_{2}\sum_{m,n}\sum_{\sigma}\nonumber\\
&&\;\;\;\;\;\;\;\;\left[(u_{in}\partial_{x}u_{im}^{*}-\partial_{x}u_{in}u_{im}^{*})\hat{c}_{m\sigma}^{+}\hat{c}_{n\sigma}\right.\nonumber\\
&&\;\;\;\;\;\;\;\;-{\sigma}(v_{in}^{*}\partial_{x}u_{im}^{*}-\partial_{x}v_{in}^{*}u_{im}^{*})\hat{c}_{m\sigma}^{+}\hat{c}_{n-\sigma}^{+}\nonumber\\
&&\;\;\;\;\;\;\;\;-{\sigma}(u_{in}\partial_{x}v_{im}-\partial_{x}u_{in}v_{im})\hat{c}_{m-\sigma}\hat{c}_{n\sigma}\nonumber\\
&&\;\;\;\;\;\;\;\;\left.+(v_{in}^{*}\partial_{x}v_{im}-\partial_{x}v_{in}^{*}v_{im})\hat{c}_{m-\sigma}\hat{c}_{n-\sigma}^{+}\right],\label{CO_1}
\end{eqnarray}
where we have replaced the sums over $j$ and $\beta$ by a single
sum other index $m$. The terms with index $m$ ($n$) corresponds to
energy $E_{1}$ ($E_{2}$). The chemical potential $\mu_{S}$ of the
superconductor is large compared to any other energy scale in the system,
thus the assumption $k_{e}=k_{h}=k_{F}$ is made here, which explains
the presence of the Fermi velocity $v_{F}$ in Eq. (\ref{CO_1}).

The calculation of current correlations is rather standard, but it
is reproduced in the Appendix so that the notations of the present
work can be understood. Compared to the calculation of the density
matrix of Sec. 3, the characterization of the scattering
properties of the setup of Fig.~\ref{fig2} now requires to include
the MZ interferometers. The magnetic fluxes which are applied on
the left ($\Phi_{A}$) and on the  right ($\Phi_{B}$)
interferometers define the Bell angles:
\begin{equation}
\varphi_{A}=2\pi\frac{\Phi_{A}}{\Phi_{0}},\;\;\;\;\;\;\;\;\varphi_{B}=2\pi\frac{\Phi_{B}}{\Phi_{0}},\label{angles}
\end{equation}

In analogy with spin entanglement of
electrons~\cite{kawabata,chtchelkatchev}, where electrons are
detected via spin filters with arbitrary polarization, the phase
angles $\varphi_{A}$ and $\varphi_{B}$ allow to rotate the
pseudo-spins with respect to the $\hat{Z}$ axis, which is defined
in a basis $|\uparrow\rangle=|u\rangle$ and
$|\downarrow\rangle=|d\rangle$. Indeed, electron wave functions
gain a factor $e^{\pm i\varphi_{A,B}/2}$ in the upper/lower arms
of the interferometers, which is equivalent to a rotation by the
angle $\varphi_{A,B}$ in the pseudo-spin basis. Next,  the use of
a symmetric beam splitter, which is described by a unitary matrix,
\begin{equation}
U=\left(
\begin{array}{ccc}
 \frac{1}{\sqrt{2}} & \frac{i}{\sqrt{2}} \\
 \frac{i}{\sqrt{2}} & \frac{1}{\sqrt{2}}
\end{array}
\right),\label{U matrix}
\end{equation}
performs a  rotation of such pseudo-spin around the $\hat{Y}$-axis
by an angle $\pi/2$. In the end, the latter measurements at the
points $A_{1}$,$A_{2}$ (left side of Fig.~\ref{fig2}) and
$B_{3}$,$B_{4}$ (right side of Fig.~\ref{fig2}) correspond to a
Stern-Gerlach type of measurement along $\hat{Z}$ axis for the
initial pseudo-spins. These transformations on both the right and
the left orbital pseudo-spins allow to perform standard Bell type
correlation measurements.

For constructing the scattering amplitudes, we use $A$, $B$, $C$,
$D$ from Eqs. (\ref{ABCD1})-(\ref{ABCD2}) and the elements of the
beam splitter scattering matrix Eq. (\ref{U matrix}). The
following convention is used for the angle signs: an electron-like
quasiparticle on the left (right) side gains a phase
$e^{+i\varphi_{A}/2}$ ($e^{-i\varphi_{B}/2}$), when it propagates
in the clockwise direction through the upper arm of the
interferometer. For anticlockwise movement the sign in the
exponent is opposite. For a hole-like quasiparticle moving
clockwise in the lower arm the sign is also opposite.

Note that this convention applies to the case where the direction
of the magnetic field is opposite in different interferometers.
This situation may seem hard to realize experimentally, however it
was chosen only for convenience: in the contrary situation (with
both fields in the same direction) one simply needs to change the
sign of an angle in (\ref{angles}).

For a given type of particle (electron or hole), the scattering
matrices have dimensions of $4\times 4$, which corresponds to the
number of leads attached to the superconductor: recall that here
we are working in the Andreev regime. In order to write the
matrices, we use the amplitudes in Eqs.
(\ref{ABCD1})-(\ref{ABCD2}), an angle
$\delta\varphi=\varphi_{A}-\varphi_{B}$ and unitary matrices for
scattering amplitudes of the beam splitters in Eq. (\ref{U
matrix}):
\begin{subequations}
\begin{equation}
\small S^{\alpha\alpha}(+E)=\left(
\begin{array}{cccc}
\frac{A}{2}+\frac{1}{2} & \frac{iA}{2}-\frac{i}{2} & \frac{B}{2}e^{-i\delta\varphi/2} & \frac{iB}{2}e^{-i\delta\varphi/2} \\
\frac{iA}{2}-\frac{i}{2} & -\frac{A}{2}-\frac{1}{2} & \frac{iB}{2}e^{-i\delta\varphi/2} & -\frac{B}{2}e^{-i\delta\varphi/2} \\
\frac{B}{2}e^{+i\delta\varphi/2} & \frac{iB}{2}e^{+i\delta\varphi/2} & \frac{A}{2}+\frac{1}{2} & \frac{iA}{2}-\frac{i}{2}\\
\frac{iB}{2}e^{+i\delta\varphi/2} &
-\frac{B}{2}e^{+i\delta\varphi/2} & \frac{iA}{2}-\frac{i}{2} &
-\frac{A}{2}-\frac{1}{2}
\end{array}
\right),\label{s_matrx1}
\end{equation}
for $\alpha\neq\beta$:
\begin{equation}
\small S^{\alpha\beta}(+E)=-\left(
\begin{array}{cccc}
\frac{C}{2} & \frac{iC}{2} & \frac{D}{2}e^{-i\delta\varphi/2} & \frac{iD}{2}e^{-i\delta\varphi/2} \\
\frac{iC}{2} & -\frac{C}{2} & \frac{iD}{2}e^{-i\delta\varphi/2} & -\frac{D}{2}e^{-i\delta\varphi/2} \\
\frac{D}{2}e^{i\delta\varphi/2} & \frac{iD}{2}e^{i\delta\varphi/2} & \frac{C}{2} & \frac{iC}{2}\\
\frac{iD}{2}e^{i\delta\varphi/2} &
-\frac{D}{2}e^{i\delta\varphi/2} & -\frac{iC}{2} & \frac{C}{2}
\end{array}
\right).\label{s_matrx2}
\end{equation}
\end{subequations}
In these matrices, the terms $\pm\frac{1}{2}$
and $+\frac{i}{2}$ in the scattering amplitudes originate from the
full reflection of a particle approaching an ideal energy filter,
as specified by Eqs. (\ref{udL1})-(\ref{udL2}), which is adjusted
not to allow the transmission of a particle with this particular
energy. The amplitudes for energies below Fermi level ($-E$) are
calculated using the property of Eq. (\ref{eh_symm}).

Note that here we did not include explicitly factors such as
$e^{\pm k_{e/h}l}$, which correspond to phases accumulated by
particles propagating through a lead of length $l$. These phases
can be included as an overall multiplication factor on the
scattering matrix in Eqs. (\ref{ABC1})-(\ref{ABC2}), depending on
the lead coordinate. Here the assumption that both arms of any
given MZ interferometer have the same length is implicit. We also
neglect the length $S$ of the injecting lead of the
superconducting fork, and we assume an ideal NS-boundary in the
pure  Andreev reflection regime with amplitudes:
$r_{SS}^{eh}=r_{SS}^{he}=e^{-i\eta}=-i$.

From the scattering amplitudes of Eqs.
(\ref{s_matrx1})-(\ref{s_matrx2}) we obtain the noise
cross-correlations, which may be separated in an excess and an
equilibrium parts:
\begin{equation}
C(x_{i},t_{i};x_{j},t_{j})=C_{ex}(x_{i},t_{i};x_{j},t_{j})+C_{eq}(x_{i},t_{i};x_{j},t_{j}),\nonumber
\end{equation}
the excess part (first term) depends on voltage and originates from
the first term in Eqs. (\ref{ABC1})-(\ref{ABC2}), the
equilibrium part(second term) is voltage independent and it occurs from the
second term (delta functions) in Eqs.
(\ref{ABC1})-(\ref{ABC2}).

It is then natural to introduce time scales which characterize the
time spacing between electron wave packets, as well as the time of
flight of particle through the wires: $\tau_{V}=\hbar/eV$,
$\tau_{ij}^{eq}=(|x_{j}|+|x_{i}|)/v_{F}$ and
$\tau_{ij}^{ex}=(|x_{j}|-|x_{i}|)/v_{F}$. In addition,
$\tau_{ij}=t_{j}-t_{i}$. So the cross-correlations at finite
temperature $T$ reads:
\begin{eqnarray}
&&C^{ex}_{ij}(\tau_{ij},\tau_{ij}^{ex},T)=\frac{2e^{2}}{h^{2}}\frac{\epsilon^{2}(1-2\epsilon-\epsilon^{2})}{8(1-\epsilon)^{4}}\nonumber\\
&&\;\;\;\;\times(1+(-1)^{i+j}\cos{(\delta\varphi-(\tau_{ij}-\tau_{ij}^{ex})/\tau_{V})})\nonumber\\
&&\;\;\;\;\times4\sin^{2}\frac{(\tau_{ij}-\tau_{ij}^{ex})}{2\tau_{V}}\frac{(\pi
T)^{2}}{\sinh^{2}(\pi T(\tau_{ij}-\tau_{ij}^{ex})/\hbar)},
\end{eqnarray}
\begin{eqnarray}
&&C^{eq}_{ij}(\tau_{ij},\tau_{ij}^{eq},T)=-\frac{2e^{2}}{h^{2}}\frac{1+\sqrt{1-2\epsilon}-(2+\sqrt{1-2\epsilon})\epsilon}{8(1-\epsilon)^{2}}\nonumber\\
&&\times\left[\frac{(\pi T)^{2}}{\sinh^{2}(\pi
T(\tau_{ij}+\tau_{ij}^{eq})/\hbar)} +\frac{(\pi
T)^{2}}{\sinh^{2}(\pi
T(\tau_{ij}-\tau_{ij}^{eq})/\hbar)}\right].\nonumber
\end{eqnarray}

For zero temperature ($k_{B}T\ll eV$), the correlators reduce to:
\begin{eqnarray}
&&C^{ex}_{ij}(\tau_{ij},\tau_{ij}^{ex},0)=\frac{2e^{4}V^{2}}{h^{2}}\frac{\epsilon^{2}(1-2\epsilon-\epsilon^{2})}{8(1-\epsilon)^{4}}
\frac{\sin^{2}\left(\frac{\tau_{ij}-\tau_{ij}^{ex}}{2\tau_{V}}\right)}{\left(\frac{\tau_{ij}-\tau_{ij}^{ex}}{2\tau_{V}}\right)^{2}}\nonumber\\
&&\;\;\;\;\;\;\;\;\times(1+(-1)^{i+j}\cos{(\delta\varphi-(\tau_{ij}-\tau_{ij}^{ex})/\tau_{V})}),\label{ex_noise}
\end{eqnarray}
\begin{eqnarray}
&&C^{eq}_{ij}(\tau_{ij},\tau_{ij}^{eq},0)=-\frac{2e^{2}}{(2\pi)^{2}}\frac{1+\sqrt{1-2\epsilon}-(2+\sqrt{1-2\epsilon})\epsilon}{8(1-\epsilon)^{2}}\nonumber\\
&&\;\;\;\;\;\;\;\;\;\;\;\;\;\;\;\;
\times\left[\frac{1}{(\tau_{ij}+\tau_{ij}^{eq})^{2}}+\frac{1}{(\tau_{ij}-\tau_{ij}^{eq})^{2}}\right].
\end{eqnarray}

Note that the equilibrium noise describes Fermi correlations
within the normal leads. At low temperatures it decays as an
inverse square with large distances from the superconducting fork
to the detectors. So assuming large distances, $x_{i},\;x_{j}\gg
v_{F}\tau_{V}$, only excess noise is relevant, and this
corresponds to the excess noise due to Cooper pairs injected from
the superconductor. The dependence on magnetic fluxes in the
excess noise occurs via $\delta\varphi=\varphi_{A}-\varphi_{B}$.
One should therefore attempt to construct a Bell type inequality
with these measurable quantities.

\section{Bell test for current correlations}

In optics, the Bell inequality test is typically expressed in
terms of correlators of number operators. This reflects the fact
that a coincidence measurement on two photons is performed. In
Refs.~\cite{lesovik_martin_blatter} and~\cite{chtchelkatchev},
these number correlators were expressed in terms of current noise
cross-correlators, which contain an irreducible contribution and a
reducible contribution. The reducible contribution is proportional
to the product of the average current in each lead. The
irreducible contribution dominates for short observation times.
Subsequently, the Bell inequality can be written in terms of
irreducible zero frequency noise cross-correlators between the
different arms of the device. A slightly different point of view
was proposed in~\cite{orbital_entanglement} and discussed
in~\cite{Entanglement4}: As long as the spacing in time between
electron wave packets $e/\langle I \rangle$ is large compared to
the size of these wave packets $\hbar/eV$, successive pairs do not
mix so that all information about entanglement is included in the
zero frequency noise correlators.

In this letter, we suggest a Bell type inequality for finite time
cross-correlations of electrical currents. It is similar in spirit
as in work of Ref.~\onlinecite{Entanglement3,Entanglement4}, where
Bell inequalities are directly expressed in terms of current
correlators, yet we shall see that our analysis is not necessarily
restricted to short times. We thus follow the derivation of
Ref.~\onlinecite{chtchelkatchev} for the Bell inequality violation
with particle numbers:
\begin{equation}
\left|E(\varphi_{a},\varphi_{b})+E(\varphi_{a},\varphi_{b}^{\prime})+E(\varphi_{a}^{\prime},\varphi_{b})-E(\varphi_{a}^{\prime},\varphi_{b}^{\prime})\right|\le
2,\label{Bell_Inq2}
\end{equation}
where
\begin{equation}
E(\varphi_{a},\varphi_{b})=\frac{\left\langle(N_{1}(\tau)-N_{2}(\tau))(N_{3}(\tau)-N_{4}(\tau))\right\rangle}{\left\langle(N_{1}(\tau)+N_{2}(\tau))(N_{3}(\tau)+N_{4}(\tau))\right\rangle},\label{Bell_term}
\end{equation}
and with $\varphi_{a}$, $\varphi_{b}$ - angles corresponding to
different magnetic fluxes in the left and the right
interferometers. $\tau$ is the duration of the measurement in
time. Particle number operators in different leads $A_{1}$-$B_{4}$
are defined from current operators:
$N_{i}(t,\tau)=\int_{t}^{t+\tau}dt^{\prime}I_{i}(t^{\prime})$. We
perform density matrix averaging and time averaging:
\begin{equation}
\left\langle
N_{i}(\tau)N_{j}(\tau)\right\rangle=\lim_{T\rightarrow\infty}\frac{1}{2T}\int_{-T}^{+T}dt\left\langle
N_{i}(t,\tau)N_{j}(t,\tau)\right\rangle_{\rho}~.
\end{equation}

We define the irreducible current cross-correlator at given
positions and times $t_{1}$ and $t_{2}$:
\begin{eqnarray}
&&C_{ij}(t_{2}-t_{1},x_{i},x_{j})=\left\langle\hat{I}_{i}(x_{i},t_{1})\hat{I}_{j}(x_{j},t_{2})\right\rangle\nonumber\\
&&\;\;\;\;\;\;\;\;\;\;\;\;\;\;\;\;
-\left\langle\hat{I}_{i}(x_{i},t_{1})\right\rangle\left\langle\hat{I}_{j}(x_{j},t_{2})\right\rangle,\label{irr_corr}
\end{eqnarray}
and we write the Bell term in the following form:
\begin{equation}
\small
E(\varphi_{a},\varphi_{b})=\frac{\int_{0}^{\tau}dt_{1}\int_{0}^{\tau}dt_{2}(C_{13}-C_{14}-C_{23}+C_{24})+\Lambda_{-}}
{\int_{0}^{\tau}dt_{1}\int_{0}^{\tau}dt_{2}(C_{13}+C_{14}+C_{23}+C_{24})+\Lambda_{+}},
\end{equation}
where $\Lambda_{\pm}$ originate from the reducible part, which is typically proportional to the average currents:
\begin{equation}
\Lambda_{\pm}=\tau^{2}\left(\left\langle
I_{1}\right\rangle\pm\left\langle
I_{2}\right\rangle\right)\left(\left\langle
I_{3}\right\rangle\pm\left\langle I_{4}\right\rangle\right).
\end{equation}
The coordinates $x_{i}$ and $x_{j}$ in Eq. (\ref{irr_corr})
correspond to positions of the detectors counted from the
superconducting fork, and $\tau$ correspond to duration of
measurements.

For the case of symmetrical beam splitters used in interferometers
and a symmetrical superconducting fork the average electrical
currents are identical: $\left\langle
I_{i}\right\rangle=\frac{e^{2}V}{h}\frac{\epsilon^{2}}{(1-\epsilon)^{2}}$,
so $\Lambda_{-}=0$ and
\begin{equation}
\small
\Lambda_{+}=4\tau^{2}\frac{e^{2}V}{h}\frac{\epsilon^{2}}{(1-\epsilon)^{2}}.
\end{equation}
In this case the Bell term reads:
\begin{equation}
\small
E(\varphi_{a},\varphi_{b})=\frac{\int_{0}^{\beta}dt_{1}\int_{0}^{\beta}dt_{2}
\frac{\sin^{2}\frac{t_{2}-t_{1}+\alpha}{2}}{(t_{2}-t_{1}+\alpha)^{2}}\cos{\left(\delta\varphi-(t_{2}-t_{1}+\alpha)\right)}}
{\int_{0}^{\beta}dt_{1}\int_{0}^{\beta}dt_{2}
\frac{\sin^{2}\frac{t_{2}-t_{1}+\alpha}{2}}{(t_{2}-t_{1}+\alpha)^{2}}+\gamma\beta^{2}},\label{main_answer}
\end{equation}
here $\alpha=\frac{x_{B}-x_{A}}{v_{F}\tau_{V}}$. For simplicity,
we consider the case where $x_{1}=x_{2}=x_{A}$ and
$x_{3}=x_{4}=x_{B}$, assuming a precision $\delta x\ll
v_{F}\tau_{V}$ - i.e. the detectors of each party $A$ and $B$ are
equidistant from the superconducting fork. Moreover,
$\beta=\tau/\tau_{V}$,
$\gamma=\frac{\epsilon^{2}}{1-2\epsilon-\epsilon^{2}}$,
$\delta\varphi=\varphi_{a}-\varphi_{b}$. So there are three
dimensionless parameters: $\alpha$ - which describes the fact that
the measurements performed by $A$ and $B$ parties are not in
coincidence; $\beta$ - the duration of these measurements,
$\gamma$ - the parameter describing the transport properties of
the superconducting fork (recall that it depends only on
$\epsilon$, which is the probability of a particle to go from the
injection lead into one of the two leads); and $\delta\varphi$ is
the difference between the Bell angles introduced in Eq.
(\ref{angles}).

In our analysis of the Bell inequality of Eq. (\ref{Bell_Inq2}) we
consider two simple cases, for both of them the Bell term in Eq.
(\ref{Bell_term}) satisfies
$E(\varphi_{a},\varphi_{b})\sim\cos{(\varphi_{a}-\varphi_{b})}$.

\begin{figure}
\centerline{\includegraphics[width=7.cm]{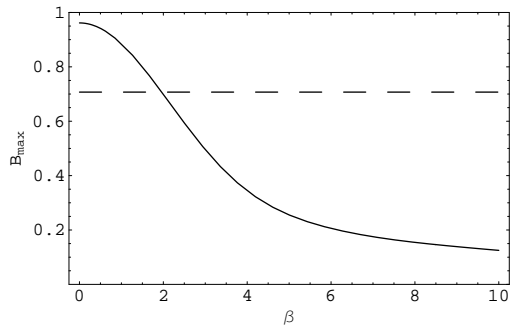}} \caption{Plot
for the maximal Bell inequality violation $B_{max}$ as a function
of the measurement time: $\beta=\tau/\tau_{V}$. The analysis is
performed in terms of current cross-correlations, the deviation
from the coincidence of measurement times
$\alpha=\frac{x_{B}-x_{A}}{v_{F}\tau_{V}}$ is taken to be zero.
The maximally possible violation is normalized to $1$, thus the
limit for Bell inequality violation is $1/\sqrt{2}$. The
transmission probability $\epsilon$ is the same as in
Fig.~\ref{plot3}.}\label{plot2}
\end{figure}

{\it i) The case of coincident measurements: $\alpha=0$.} On
Fig.~\ref{plot2} we show the resulting dependence of the maximal
Bell inequality violation $B_{max}$ (which is the maximum of the
left side of the Bell inequality over the Bell angles) over the
measurement time $\tau$. From Fig.~\ref{plot2} it is clear, that
for the case of superconducting fork with very bad transmission
($\epsilon\to 0$, $\gamma\to 0$), the violation of Bell inequality
is possible only for $\tau<2\tau_{V}$. We again stress that long
measurement times destroy the interference between trajectories
with different energies, so that the Bell inequality cannot be
violated.

In our analysis, the maximal violation depends on the transmission
properties of the superconducting fork: the closer $\gamma$ to
zero the better the violation. The Bell inequality in Eq.
(\ref{Bell_Inq2}) may be violated maximally (with $\beta\to 0$)
for $\epsilon\to 0$, and it is not violated at all for
$\epsilon\ge \epsilon_{max}=(3-\sqrt{2})/7\approx 0.22$ [this
corresponds to: $\gamma\ge\gamma_{max}=(\sqrt{2}-1)/4$]. This fact
may by explained in the following way: the irreducible
cross-correlations vanishes when $\gamma\to \pm\infty$ (See
Ref.~\onlinecite{Noise3}). Indeed, the term of Eq.
(\ref{ex_noise}) equals to zero for
$\epsilon_{0}=\sqrt{2}-1\approx 0.42$.

{\it ii) The case of short time measurements: $\beta\rightarrow
0$.} Here we vary two parameters: $\epsilon$ - the transparency of
the superconducting fork, which defines the parameter
$\gamma=\frac{\epsilon^{2}}{1-2\epsilon-\epsilon^{2}}$, and
$\alpha=\frac{x_{B}-x_{A}}{v_{F}\tau_{V}}$ - which characterizes
the lack of coincidence of the measurements by the two parties $A$
and $B$. Again, the best violation occurs for $\alpha\to 0$, and
there is a dependence of maximal violation on transmission
properties of the fork, which are described by $\gamma$. The Bell
inequality in this case is the following:
\begin{equation}
B_{max}=\left|\frac{\sin^{2}{\frac{\alpha}{2}}}{\sin^{2}{\frac{\alpha}{2}}+\gamma\alpha^{2}}\right|\le\frac{1}{\sqrt{2}},\label{Bell_Inq3}
\end{equation}
where we have chosen the Bell angles:
$\delta\varphi_{ab}=\delta\varphi_{ab^{\prime}}=\delta\varphi_{a^{\prime}b}=\pi/4-\alpha$,
$\delta\varphi_{a^{\prime}b^{\prime}}=3\pi/4-\alpha$, for which
the violation is maximal. The dependence of these angles on
$\alpha$ is due to the presence of $\alpha$ under the cosine in
Eq. (\ref{main_answer}).

In Fig.~\ref{plot3} we plot the dependence of the left hand side
of the Bell inequality on $\alpha$, for a certain $\gamma=0.001$.
There is again a threshold in the transparency
$\epsilon_{max}=0.22$, below which the violation is possible. For
a transmission probability below the threshold $\epsilon_{max}$
the Bell inequality in Eq. (\ref{Bell_Inq3}) can be violated for:
$\alpha\le\tau_{V}\sqrt{(\sqrt{2}-1)/\gamma}\approx
0.6\tau_{V}/\epsilon$, while it remains inviolated in narrow
regions separated by intervals: $\frac{h}{eV}$.

To compare the entanglement measure and the maximal Bell
inequality violation, we plot in the same Fig.~\ref{plot3} the
concurrency ${\bf C}$ as well as the function  $\sqrt{\frac{1+{\bf
C}^{2}}{2}}$. It is known~\cite{gisin} that for mixed states
$B_{max}$ is confined between these two values (for the case of
pure states $B_{max}\equiv\sqrt{\frac{1+{\bf C}^{2}}{2}}$), which
is clearly seen from the plots of Fig.~\ref{plot3}. According to
our results, there may occur situations, where the state on the
detectors is entangled (${\bf C}\ne 0$), but there is no Bell
inequality violation ($B_{max}<1/\sqrt{2}$).

\begin{figure}
\centerline{\includegraphics[width=7.cm,height=5.cm]{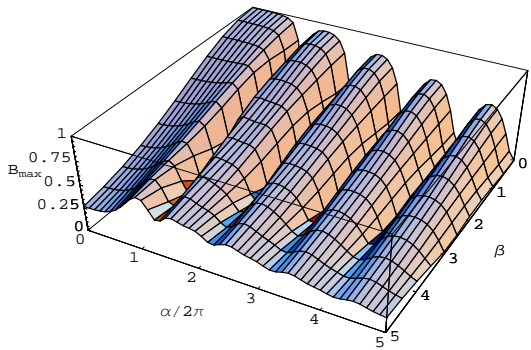}}
\caption{(Color online) Plot for the maximal Bell inequality
violation $B_{max}$ depending both on the deviation from
coincidence of current measurements:
$\alpha=\frac{x_{B}-x_{A}}{v_{F}\tau_{V}}$ and the measurement
time $\beta\tau\ll\hbar/eV$. The maximal value is normalized to
unity. The transmission probability is taken to be:
$\gamma\approx\epsilon^{2}=0.001$. Violation above the critical
value of $1/\sqrt{2}$ occurs only at separate peak
locations.}\label{plot4}
\end{figure}

\begin{figure}
\centerline{\includegraphics[width=7.cm,height=5.cm]{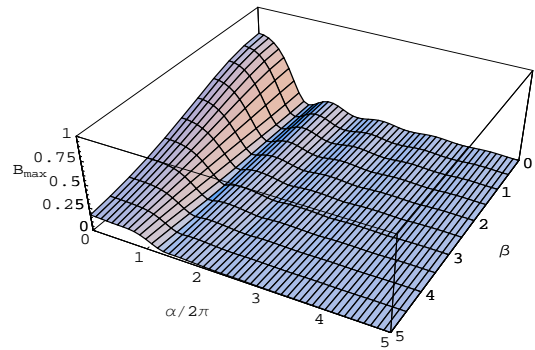}}
\caption{(Color online) Same as Fig.~\ref{plot4}, except that the
transmission probability is taken to be critical:
$\gamma=\gamma_{max}\approx 0.1$. The whole surface lies below the
critical value $1/\sqrt{2}$, so there is no violation of the Bell
inequality.}\label{plot5}
\end{figure}

From these two examples of {\it i)} $\alpha=0$ and {\it ii)}
$\beta=0$ we conclude that the violation of Bell inequality occurs
for small measurement times: $\tau<2\tau_{V}$ and for measurements
which are close to being coincident:
$|x_{A}-x_{B}|/v_{F}<0.6\tau_{V}/\epsilon$. We performed numerical
integrations of Eq. (\ref{main_answer}) for arbitrary values of
$\{\alpha,\beta,\gamma\}$. The results for $\gamma=0.001$ and
$\gamma=\gamma_{max}\approx 0.1$ are shown on two-dimensional
plots of Fig.~\ref{plot4} and of Fig.~\ref{plot5}. Note that the
plot of Fig.~\ref{plot3} represent a slice at $\alpha=0$ of
Fig.~\ref{plot4}, while Fig.~\ref{plot2} represents a slice at
$\beta=0$ of the same figure. One notices that the oscillations of
$B_{max}$ are damped as $\alpha$ is increased, and, as mentioned
above that violation is less likely when $\beta$ is increased
because the time window is too large. Such oscillations in the
Bell parameter were detected previously in the case of normal
metal forks~\cite{Entanglement4}. It is obvious from
Fig.~\ref{plot5} that, when the transparency of the
superconducting fork is larger than the critical threshold, there
is no violation of the Bell inequalities, although oscillations
are still noticeable.

\section{Conclusion.}

We have proposed a setup measuring energy entanglement for pairs
of quasiparticles originating from single Cooper pairs in a
superconductor. The use of energy filters allows to separate the
particles according to their energy and to obtain orbitally
entangled pairs of electrons, which are then analyzed. This
electronic setup and its detection apparatus is an analog of the
momentum-phase entanglement performed in quantum
optics~\cite{rarity}. a) Two electrons from a Cooper pair
correspond to the two photons generated by down-conversion, which
have different wave lengths. Indeed, Andreev reflection, or the
emission of a Cooper pair, also involves two electrons with
different energies. b) In optics the spatial separation of the
photon beams follows automatically from the down conversion
process. Here, we used energy filters to perform such separation
and considered short time measurements to allow interference of
particles with slightly different energies. It is because the
interference occurs between different energies that the ``usual''
zero-frequency analysis is not applicable here. This is why we
developed a Bell inequality test for finite-times
cross-correlations of electrical currents.

First, we proposed a density matrix analysis for quasiparticles
emitted from a superconductor and separated in energy. This allows
to calculate the concurrence, one of the most widely used
entanglement measures. Second, we showed how energy entanglement
can be transformed into orbital entanglement and how it then can
be tested via electrical current cross-correlations. For such
cross-correlations, we constructed a Bell type inequality and
showed what set of parameters lead to its violation.

The noise cross-correlations between the left and the right
reservoirs were derived in the context of the scattering theory
for normal metal-superconductor hybrid circuits. Indeed, in
stationary quantum transport these correlations enter explicitly
the Bell inequality test. Here the manifestation of entanglement
is found by adjusting the phases accumulated by electrons and
holes in the conductor leading to the beam splitter (one can
change magnetic fluxes within MZ interferometers by varying the
magnetic field through them or by varying their area, like it was
done in experiments with a single
interferometer~\cite{MZ_for_electrons}). For a device consisting
of perfect elements (filters, beam splitters,...), maximal Bell
violation is obtained for short time of measurements:
$\tau<2\tau_{V}$ and small deviations from coincident measurements
between the left and the right reservoirs:
$|x_{A}-x_{B}|/v_{F}<0.6\tau_{V}/\epsilon$ - for relatively small
transparency $\epsilon$ of the superconducting fork. We found that
the violation of the Bell inequality depends on transmission
properties of the fork, and it occurs for:
$\epsilon<\epsilon_{max}\approx 0.22$.

A clear issue is to inquire whether such energy entanglement can
be detected in experiments. Our setup for generating entangled
electron pairs has many components, each of which in principle can
be built for instance using gated semiconductor hetero-structures.
The design of a controllable normal metal-superconducting fork
represents a challenge but, recently, work in this direction has
been promising~\cite{schonenberger}. All leads are assumed one
dimensional, and are assumed to have little or no backscattering
as, for instance, in the experiments of Ref.~\cite{auslaender}.
The energy filtering fork can be constructed from an ordinary fork
with coherent quantum dots at two ends of it. The dots are
adjusted to have transmission peaks symmetrically above and below
the Fermi level. For better filtering these dots must have levels
with an energy spacing larger than the bias voltage.

Because of the complexity of this device, there are other
complications which should be considered.  The setup is composed
of two MZ interferometers, which are both subject to
dephasing~\cite{marquard}. The fluctuations in the gate voltages
could trigger fluctuations in the path length, which consequently
would add an uncertainty to the phase of electrons and holes,
which enter the noise correlation expression of Eq.
(\ref{ex_noise}). As long as the phase fluctuations are controlled
(small compared to a phase angle $\pi$, meaning that the
fluctuations in path length are below the Fermi wave length
$\lambda_F$), the Bell inequality will continue to be violated.
Furthermore, a considerable precision is required in the building
of such an interferometer: the two arms need to have equal length,
up to a distance $v_F\tau_V>\lambda_F$.

In conclusion, we have proposed the first energy entanglement
setup for testing the non-locality properties of electrons in
nanostructures. With the previous work on spin entanglement using
NS junctions, this latest work emphasizes the analogy with quantum
optics. Our main goal here has been to show that energy
entanglement can in principle be analyzed (concurrence) or
detected (noise cross-correlation analysis) in this somewhat ideal
device, using concepts and circuitry borrowed from mesoscopic
physics.

We thank Andrey Lebedev and Pavel Ostrovsky for fruitful
discussions and helpful comments. K.B. acknowledges support by the
ENS-Landau exchange program. G.L. and K.B. acknowledges support
from the Russian Foundation for Basic Research, grant
06-02-17086-a. T.M. acknowledges support from an AC Nanosciences
and from the ACI Jeunes Chercheurs (Lefloch) funding programs of
CNRS.

\begin{widetext}
\appendix
\section{Density matrix}

Starting from Eqs. (\ref{for_appendix1})-(\ref{for_appendix2}) we
average two anihilation operators products as:
\begin{equation}
\left\langle
\hat{c}_{m\sigma}^{+}(E_{1})\hat{c}_{n\sigma}(E_{2})\right\rangle=f_{m}\delta_{mn}\delta(E_{1}-E_{2})~,
\end{equation}
where $f_{m}=f_{FD}(E{\mp}eV)$ for electrons and holes, with
$f_{FD}(E)$ the Fermi-Dirac distribution. And averages of four
operators are calculated using Wick's theorem:
\begin{eqnarray}
&&\;\;\;\;\;\;\;\;\;\;\;\;\;\;\;\;
\left\langle\hat{c}_{k\sigma_{1}^{\prime}}^{\dagger}(E_{1},t_{i^{\prime}})\hat{c}_{l\sigma_{2}^{\prime}}^{\dagger}(E_{2},t_{j^{\prime}})
\hat{c}_{m\sigma_{2}}(E_{3},t_{j})\hat{c}_{n\sigma_{1}}(E_{4},t_{i})\right\rangle=\nonumber\\
&&=f_{k}(E_{1})f_{l}(E_{2})\delta_{\sigma_{1}^{\prime}\sigma_{1}}\delta_{\sigma_{2}^{\prime}\sigma_{2}}\delta_{kn}\delta_{lm}
\delta(E_{1}-E_{4})\delta(E_{2}-E_{3})e^{-iE_{1}(t_{i}-t_{i^{\prime}})/\hbar}e^{+iE_{2}(t_{j}-t_{j^{\prime}})/\hbar}\nonumber\\
&&-f_{k}(E_{1})f_{l}(E_{2})\delta_{\sigma_{1}^{\prime}\sigma_{2}}\delta_{\sigma_{2}^{\prime}\sigma_{1}}\delta_{km}\delta_{ln}
\delta(E_{1}-E_{3})\delta(E_{2}-E_{4})e^{-iE_{1}(t_{j}-t_{i^{\prime}})/\hbar}e^{+iE_{2}(t_{j^{\prime}}-t_{i})/\hbar}.\label{Wick}
\end{eqnarray}
The resulting calculation leads to appearance of three terms with
different spin symmetry:
\begin{eqnarray}
&&\;\;\;\;\;\;\;\;g(\vec{\sigma},\vec{x})=\frac{1}{(hv_{F})^{2}}\sum\limits_{k,l}\int_{0}^{+\infty}\int_{0}^{+\infty}dE_{1}dE_{4}\label{DM1}\\
&&\left\{\delta_{\sigma_{1}\sigma_{1}^{\prime}}\delta_{\sigma_{2}\sigma_{2}^{\prime}}
\left[(u_{i^{\prime}k}^{*}(x_{i^{\prime}})u_{ik}(x_{i}))(u_{jl}(x_{j})u_{j^{\prime}l}^{*}(x_{j^{\prime}}))e^{-iE_{1}(t_{i}-t_{i^{\prime}})}e^{+iE_{2}(t_{j^{\prime}}-t_{j})}f_{k}(E_{1})f_{l}(E_{2})\right.\right.\\
&&\;\;\;\;\;\;\;\;\;\;\;\;\;\;\;\;\;+(u_{i^{\prime}k}^{*}(x_{i^{\prime}})u_{ik}(x_{i}))(v_{jl}^{*}(x_{j})v_{j^{\prime}l}(x_{j^{\prime}}))e^{-iE_{1}(t_{i}-t_{i^{\prime}})}e^{-iE_{2}(t_{j^{\prime}}-t_{j})}f_{k}(E_{1})(1-f_{l}(E_{2}))\\
&&\;\;\;\;\;\;\;\;\;\;\;\;\;\;\;\;\;+(v_{i^{\prime}k}(x_{i^{\prime}})v_{ik}^{*}(x_{i}))(u_{jl}(x_{j})u_{j^{\prime}l}^{*}(x_{j^{\prime}}))e^{+iE_{1}(t_{i}-t_{i^{\prime}})}e^{+iE_{2}(t_{j^{\prime}}-t_{j})}(1-f_{k}(E_{1}))f_{l}(E_{2})\\
&&\left.\;\;\;\;\;\;\;\;\;\;\;\;\;\;\;\;\;+(v_{i^{\prime}k}(x_{i^{\prime}})v_{ik}^{*}(x_{i}))(v_{jl}^{*}(x_{j})v_{j^{\prime}l}(x_{j^{\prime}}))e^{+iE_{1}(t_{i}-t_{i^{\prime}})}e^{-iE_{2}(t_{j^{\prime}}-t_{j})}(1-f_{k}(E_{1}))(1-f_{l}(E_{2}))\right]
\end{eqnarray}
\begin{eqnarray}
&&-\delta_{\sigma_{1}\sigma_{2}^{\prime}}\delta_{\sigma_{2}\sigma_{1}^{\prime}}
\left[ (u_{i^{\prime}k}^{*}(x_{i^{\prime}})u_{jk}(x_{j}))(u_{il}(x_{i})u_{j^{\prime}l}^{*}(x_{j^{\prime}}))e^{-iE_{1}(t_{j}-t_{i^{\prime}})}e^{+iE_{2}(t_{j^{\prime}}-t_{i})}f_{k}(E_{1})f_{l}(E_{2})\right.\\
&&\;\;\;\;\;\;\;\;\;\;\;\;\;\;\;\;\;+(u_{i^{\prime}k}^{*}(x_{i^{\prime}})u_{jk}(x_{j}))(v_{il}^{*}(x_{i})v_{j^{\prime}l}(x_{j^{\prime}}))e^{-iE_{1}(t_{j}-t_{i^{\prime}})}e^{-iE_{2}(t_{j^{\prime}}-t_{i})}(1-f_{l}(E_{2}))\\
&&\;\;\;\;\;\;\;\;\;\;\;\;\;\;\;\;\;+(v_{i^{\prime}k}(x_{i^{\prime}})v_{jk}^{*}(x_{j}))(u_{il}(x_{i})u_{j^{\prime}l}^{*}(x_{j^{\prime}}))e^{+iE_{1}(t_{j}-t_{i^{\prime}})}e^{+iE_{2}(t_{j^{\prime}}-t_{i})}(1-f_{k}(E_{1}))f_{l}(E_{2})\\
&&\left.\;\;\;\;\;\;\;\;\;\;\;\;\;\;\;\;\;+(v_{i^{\prime}k}(x_{i^{\prime}})v_{jk}^{*}(x_{j}))(v_{il}^{*}(x_{i})v_{j^{\prime}l}(x_{j^{\prime}}))e^{+iE_{1}(t_{j}-t_{i^{\prime}})}e^{-iE_{2}(t_{j^{\prime}}-t_{i})}(1-f_{k}(E_{1}))(1-f_{l}(E_{2}))\right]\\
&&+I_{\sigma_{1}\sigma_{2}}I_{\sigma_{1}^{\prime}\sigma_{2}^{\prime}}
\left[ (u_{i^{\prime}k}^{*}(x_{i^{\prime}})v_{j^{\prime}k}(x_{j^{\prime}}))(u_{il}(x_{i})v_{jl}^{*}(x_{j}))e^{-iE_{1}(t_{j^{\prime}}-t_{i^{\prime}})}e^{+iE_{2}(t_{j}-t_{i})}f_{k}(E_{1})f_{l}(E_{2})\right.\\
&&\;\;\;\;\;\;\;\;\;\;\;\;\;\;\;\;\;-(u_{i^{\prime}k}^{*}(x_{i^{\prime}})v_{j^{\prime}k}(x_{j^{\prime}}))(v_{il}^{*}(x_{i})u_{jl}(x_{j}))e^{-iE_{1}(t_{j^{\prime}}-t_{i^{\prime}})}e^{-iE_{2}(t_{j}-t_{i})}f_{k}(E_{1})(1-f_{l}(E_{2}))\\
&&\;\;\;\;\;\;\;\;\;\;\;\;\;\;\;\;\;-(v_{i^{\prime}k}(x_{i^{\prime}})u_{j^{\prime}k}^{*}(x_{j^{\prime}}))(u_{il}(x_{i})v_{jl}^{*}(x_{j}))e^{+iE_{1}(t_{j^{\prime}}-t_{i^{\prime}})}e^{+iE_{2}(t_{j}-t_{i})}(1-f_{k}(E_{1}))f_{l}(E_{2})\\
&&\left.\left.\;\;\;\;\;\;\;\;\;\;\;\;\;\;\;\;\;+(v_{i^{\prime}k}(x_{i^{\prime}})u_{j^{\prime}k}^{*}(x_{j^{\prime}}))(v_{il}^{*}(x_{i})u_{jl}(x_{j}))e^{+iE_{1}(t_{j^{\prime}}-t_{i^{\prime}})}e^{-iE_{2}(t_{j}-t_{i})}(1-f_{k}(E_{1}))(1-f_{l}(E_{2}))\right]\right\},\label{DM1_end}
\end{eqnarray}
where we have used the matix notation ($\sigma_i=\pm 1$):
$I_{\sigma_{1}\sigma_{2}}=i\hat{\tau}_y$ and $\hat{\tau}_y$ is a
Pauli matrix.

We consider the Andreev reflection at the NS-boundary to be ideal (no normal reflection),
and the reflection amplitude reads:
$r_{A}=e^{-i\arccos{\frac{E}{\Delta}}}$. According to the
properties of Bogolubov's equations the scattering amplitudes must
satisfy the equations:
\begin{equation}
S_{ij}^{hh}(E)=S_{ij}^{ee*}(-E),\;\;\;\;S_{ij}^{he}(E)=-S_{ij}^{eh*}(-E),\label{eh_symm}
\end{equation}
from this, one can derive the property:
\begin{equation}
v_{ij\beta}(x,E)=(-1)^{\delta_{e\beta}}u_{ij\overline{\beta}}^{*}(x,-E),\;\;\;\;
u_{ij\beta}(x,E)=(-1)^{\delta_{h\beta}}v_{ij\overline{\beta}}^{*}(x,-E),
\end{equation}
and using the identity $f_{h}(E)=1-f_{e}(-E)$ one can simplify
Eqs. (\ref{DM1})-(\ref{DM1_end}):
\begin{eqnarray}
&&g(\vec{\sigma},\vec{x})=\delta_{\sigma_{1}\sigma_{1}^{\prime}}\delta_{\sigma_{2}\sigma_{2}^{\prime}}G_{i^{\prime}i}G_{jj^{\prime}}^{*}-
\delta_{\sigma_{1}\sigma_{2}^{\prime}}\delta_{\sigma_{2}\sigma_{1}^{\prime}}G_{i^{\prime}j}G_{ij^{\prime}}^{*}+
I_{\sigma_{1}\sigma_{2}}I_{\sigma_{1}^{\prime}\sigma_{2}^{\prime}}F_{i^{\prime}j^{\prime}}F_{ij}^{*},
\end{eqnarray}
where
\begin{subequations}
\begin{eqnarray}
&&G_{ij}(x_{i},x_{j},t_{j}-t_{i})=\frac{1}{hv_{F}}\sum\limits_{k}\int_{-\infty}^{+\infty}dE f_{k}(E)(u_{ik}^{*}(x_{i})u_{jk}(x_{j}))e^{-iE(t_{j}-t_{i})},\\
&&F_{ij}(x_{i},x_{j},t_{j}-t_{i})=\frac{1}{hv_{F}}\sum\limits_{k}\int_{-\infty}^{+\infty}dE
f_{k}(E)(u_{ik}^{*}(x_{i})v_{jk}(x_{j}))e^{-iE(t_{j}-t_{i})}.
\end{eqnarray}
\end{subequations}

\section{Current cross-correlations}

It is convenient to define the following matrix elements:
\begin{subequations}
\begin{eqnarray}
&&A_{imin}(E,E^{\prime},x)=u_{in}(E^{\prime},x)\partial_{x}u_{im}^{*}(E,x)-\partial_{x}u_{in}(E^{\prime},x)u_{im}^{*}(E,x),\label{ME_1}\\
&&B_{imin}(E,E^{\prime},x)=v_{in}^{*}(E^{\prime},x)\partial_{x}v_{im}(E,x)-\partial_{x}v_{in}^{*}(E^{\prime},x)v_{im}(E,t),\\
&&C_{imin}(E,E^{\prime},x)=u_{in}(E^{\prime},x)\partial_{x}v_{im}(E,x)-\partial_{x}u_{in}(E^{\prime},x)v_{im}(E,x).\label{ME_2}
\end{eqnarray}
\end{subequations}

These matrix elements have a useful symmetry:
\begin{subequations}
\begin{eqnarray}
&&A_{imin}(E,E^{\prime},x)=-A_{inim}^{*}(E^{\prime},E,x),\\
&&B_{imin}(E,E^{\prime},x)=-B_{inim}^{*}(E^{\prime},E,x).
\end{eqnarray}
\end{subequations}

The answer for an average current is the following:
\begin{equation}
\left\langle\hat{I}_{i}(x,t)\right\rangle=\frac{ie}{2\pi
mv_{F}}\int_{0}^{+\infty}dE\sum_{m}\left[
A_{imim}(E,E,x)f_{m}+B_{imim}(E,E,x)(1-f_{m})\right].\label{av_curr}
\end{equation}

The calculation of irreducible cross-correlation gives:
\begin{eqnarray}
&&\left\langle\hat{I}_{i}(x_{i},t_{i})\hat{I}_{j}(x_{j},t_{j})\right\rangle-\left\langle\hat{I}_{i}(x_{i},t_{i})\right\rangle\left\langle\hat{I}_{j}(x_{j},t_{j})\right\rangle=\frac{e^{2}\hbar^{2}}{2m^{2}v_{F}^{2}}\frac{1}{(2\pi\hbar)^{2}}
\int_{0}^{+\infty}dE\int_{0}^{+\infty}dE^{\prime}\sum_{m,n}\label{corr_6}\\
&&\left[
f_{m}(E)(1-f_{n}(E^{\prime}))e^{-i(E-E^{\prime})(t_{j}-t_{i})/\hbar}
\left[
A_{imin}(E,E^{\prime},x_{i})+B_{imin}^{*}(E,E^{\prime},x_{i})\right]
\left[ A_{jmjn}^{*}(E,E^{\prime},x_{j})+B_{jmjn}(E,E^{\prime},x_{j})\right]\right.\nonumber\\
&&+f_{m}(E)f_{n}(E^{\prime})e^{-i(E+E^{\prime})(t_{j}-t_{i})/\hbar}
C_{inim}^{*}(E^{\prime},E,x_{i})\left[ C_{jnjm}(E^{\prime},E,x_{j})+C_{jmjn}(E,E^{\prime},x_{j})\right]\nonumber\\
&&\left.+(1-f_{m}(E))(1-f_{n}(E^{\prime}))e^{+i(E+E^{\prime})(t_{j}-t_{i})/\hbar}
C_{imin}(E,E^{\prime},x_{i})\left[
C_{jmjn}^{*}(E,E^{\prime},x_{j})+C_{jnjm}^{*}(E^{\prime},E,x_{j})\right]\right].\nonumber
\end{eqnarray}
The averages of products for two and four anihilation operators
were calculated using Wick's theorem in Eq. (\ref{Wick}).

In Eq. (\ref{corr_6}) all terms like in Eqs.
(\ref{ME_1})-(\ref{ME_1}) may be written in terms of the
scattering matrix elements $s_{ij}^{\alpha\beta}$:
\begin{subequations}
\begin{eqnarray}
&A_{ik\alpha,il\beta}(E,E^{\prime},x)&=2ik_{F}s_{ik}^{e\alpha*}(E)s_{il}^{e\beta}(E^{\prime})e^{+i(E-E^{\prime})\frac{x}{v_{F}\hbar}}
-2ik_{F}\delta_{\alpha e}\delta_{\beta e}\delta_{ik}\delta_{il}e^{-i(E-E^{\prime})\frac{x}{v_{F}\hbar}}\label{ABC1}\\
&B_{ik\alpha,il\beta}(E,E^{\prime},x)&=2ik_{F}s_{ik}^{h\alpha}(E)s_{il}^{h\beta*}(E^{\prime})e^{-i(E-E^{\prime})\frac{x}{v_{F}\hbar}}
-2ik_{F}\delta_{\alpha h}\delta_{\beta h}\delta_{ik}\delta_{il}e^{+i(E-E^{\prime})\frac{x}{v_{F}\hbar}}\\
&C_{ik\alpha,il\beta}(E,E^{\prime},x)&=2ik_{F}s_{ik}^{h\alpha}(E)s_{il}^{e\beta}(E^{\prime})e^{-i(E+E^{\prime})\frac{x}{v_{F}\hbar}}
-2ik_{F}\delta_{\alpha h}\delta_{\beta
e}\delta_{ik}\delta_{il}e^{+i(E+E^{\prime})\frac{x}{v_{F}\hbar}},\label{ABC2}
\end{eqnarray}
\end{subequations}
here we have linearized the $k$-vectors in
energy, thus we are not interested in the quadratic behavior of
the spectrum near Fermi level and hence we neglect the dispersion
of electron and hole wave packets.

\end{widetext}

\end{document}